\documentclass[sigconf]{acmart}

\AtBeginDocument{%
  }

\usepackage{multirow}
\begin{document}

\title{Using Helium Balloon Flying Drones for Introductory CS Education}

\author{Stanley Cao}
\affiliation{%
  \institution{Stanford University}
  \city{Stanford}
  \state{CA}
  \country{USA}
}

\author{Christopher Gregg}
\affiliation{%
  \institution{Stanford University}
  \city{Stanford}
  \state{CA}
  \country{USA}}








\begin{abstract}
 In the rapidly evolving field of computer science education, engaging and accessible teaching methods are crucial for fostering interest among a diverse student body. This paper presents a new drone-based curriculum designed to introduce high school students to fundamental programming concepts through hands-on learning with inexpensive helium-balloon drones. Our approach uses a custom web-based coding environment, enabling students to experiment with Python without the need for extensive software setup. The curriculum covers basic programming skills and offers advanced projects, such as mobile app development for drone control and inter-drone communication for flocking tasks. By integrating physical computing with algorithmic thinking, we aim to enhance student engagement, critical thinking, and problem-solving abilities. Preliminary implementation plans and potential challenges are discussed. Also discussed is future work focusing on evaluating the curriculum's impact on student learning outcomes and interest in STEM fields.
\end{abstract}

\maketitle

\section{Introduction}
In the rapidly evolving field of computer science education, novel approaches to teaching fundamental concepts are crucial for engaging a diverse student body. Given the growing demand for a computing-skilled workforce, it is essential to adapt educational methods to capture the interest of a broader audience than what current computing education typically targets. Engaging educational experiences have been shown to have a positive impact on learning outcomes and examination performance, especially within computing education \cite{active_learning, ramasamy, Freeman2014ActiveLI, gan_kok}. Moreover, physical computing devices have been shown to correlate with increased student motivation when students are studying computer science \cite{apiola_creativity, buechley_engagement}. 

While many courses in higher education explore the latest research trends as tools for engagement, this approach falls short in introductory classes where it is needed most --- to appeal to a broader audience of students with limited experience in computer science. 

Our research project explores the viability of using inexpensive helium-balloon drones as a teaching tool for introducing computer science principles to high school students, and this paper documents the progress that we have made. We have engineered the drones, drawing inspiration from blimps, so that each one features a helium-filled balloon equipped with compact propellers and a microprocessor. The drones are expected to be programmed wirelessly in Python through a web-based interface.

Our newly proposed drone-based curriculum aims to cater to students with varying levels of programming experience. For students with little prior programming knowledge, the curriculum introduces the basics of a Python-based API that enables interaction with drones. Our aim is to not only teach fundamental programming concepts such as function calling and procedural coding, but to also engage students with hands-on, interactive learning experiences centered around the intersection of many fields (e.g., robotics, computer science, and electrical engineering). Advanced students are challenged with more complex projects, such as developing a mobile app for drone control. This app serves as both a practical tool for navigating drones and a creative outlet that allows students to experiment with real-world applications of programming.

A highlight of the curriculum is the final creative project, for which we propose several ideas. The first project allows students to focus on the individual control of drones, culminating in a competitive time trial through an obstacle course. The second, more sophisticated project involves "flocking" behavior, where drones must be programmed to work collaboratively towards a common goal, utilizing onboard sensors to complete complex tasks.

The curriculum will be delivered to high school students with an interest in computer science and robotics, spanning a one to two-week period. Following the implementation, we plan to compile an experience report to document the students' learning journey, their engagement with the curriculum, and the effectiveness of drones as educational tools. The remainder of this paper outlines the structure of a drone-based curriculum, with the goal of assessing whether such a curriculum can serve as an effective method for teaching computer science principles to a diverse student population.




\section{Background}
Various studies have incorporated physical electronics into the computer science curriculum. Tangible devices including Arduino microcontrollers, Android phones, and Sifteo Cubes have been integrated into a curriculum encouraging students' exposure to a wide range of physical computing devices \cite{Goadrich}. The combination of a maker-space environment with concepts in programming and computer science theory (e.g., finite automata) has been explored \cite{chamberlain}. \citet{bender} have proposed a senior-level capstone project using the Arduino platform to construct autonomous robots and contribute to faculty research.

Particularly noteworthy is the prior work that has investigated drones and robots for the teaching of introductory programming principles. 
\citet{anderson} show that first-year students benefit from robotics-based programming exercises by learning to use sensors, model data, and design algorithms, all of which are experiential and team-based, creating memorable teaching moments. Moreover, drones have served as an excellent teaching tool for basic software development and engagement with advanced concepts that span multiple computing specialties, such as autonomous flight, computer vision, flight emergence exception handling, and even ethical concerns with Unmanned Aerial Vehicles (UAV) \cite{albina}.

This paper further explores the use of drones in introductory computer science education, with particular emphasis on scalability, and inclusiveness. Our engineered drones do not involve the use of purchased quad-copters as is the case for \citet{albina}; rather a blimp-inspired design is used to create inexpensive drones that are affordable and easy to integrate into current high-school curricula. We propose a computer science curriculum that is engaging and accessible for students with limited programming experience. Our goal is to make computer science education more inclusive and to equip students with skills that are valuable across various disciplines.

\section{Hardware}
\begin{figure}[h]
    \centering
    \includegraphics[width=0.5\textwidth]{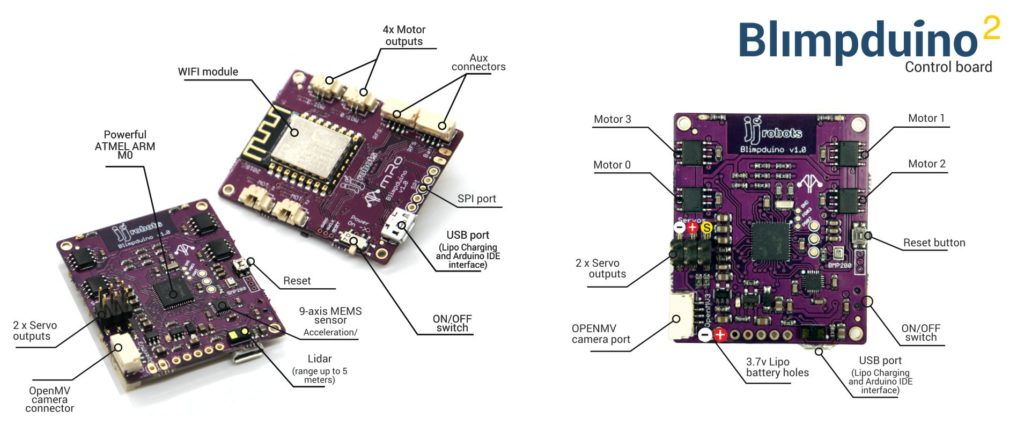}
    \caption{Blimpduino chip details. \cite{blimpduino_assembly_guide}}
    \label{fig:PCB}
\end{figure}

Our drones are inspired by the Blimpduino 2 design created by JJRobots, a company that provides instructions and hardware for robotic projects \cite{blimpduino2, jjrobots_home}. We used an open source PCB design similar to the Blimpduino 2 chip shown in Figure \ref{fig:PCB}, though we made a few minor substitutions for hardware parts that were no longer available. 

Note that the computer chip contains a main CPU and a WiFi module, which are used to receive student-written code wirelessly from a computer for compilation and execution on the computer chip.

\begin{figure}[h]
    \centering
    \includegraphics[width=0.5\textwidth]{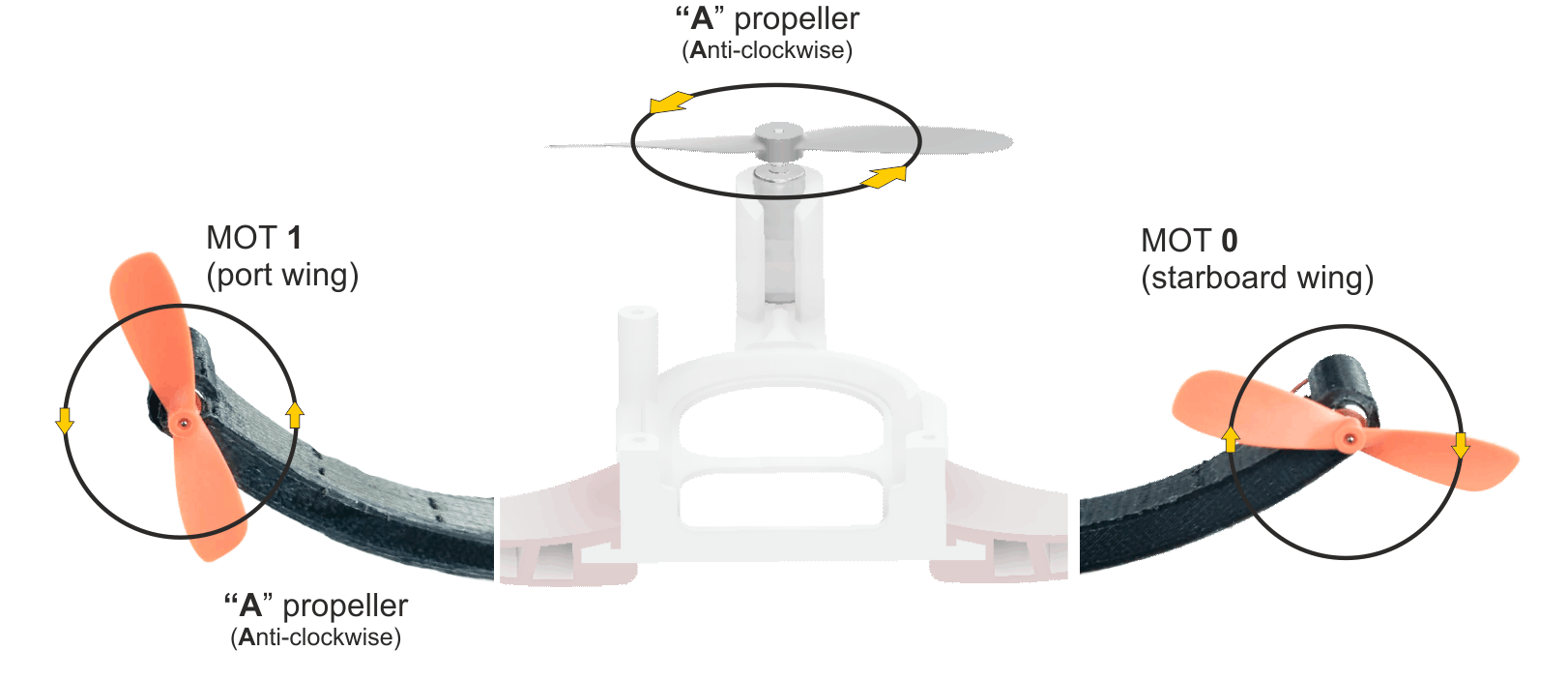}
    \caption{Computer chip encasing with propellers attached \cite{blimpduino_assembly_guide}}
    \label{fig:propellers}
\end{figure}

Figure \ref{fig:propellers} shows the computer chip encasing that we created using a 3D printer. Our device will have three motors for controlling altitude, yaw, and lateral (i.e., forward-backward) movement.

\begin{figure}[h]
    \centering
    \includegraphics[width=0.5\textwidth]{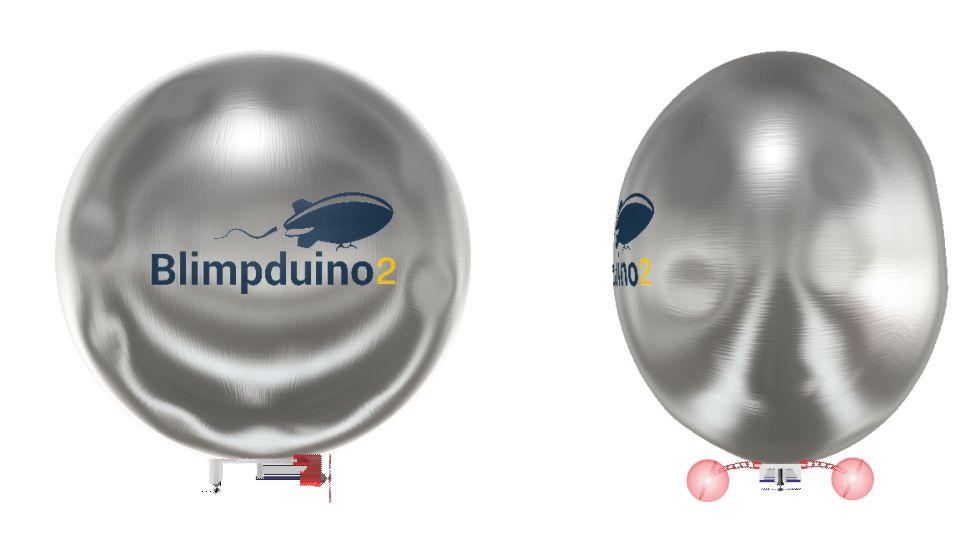}
    \caption{Illustration of complete Blimpduino drone \cite{blimpduino_assembly_guide}}
    \label{fig:full_blimp}
\end{figure}

Figure \ref{fig:full_blimp} illustrates how the chip encasing is attached to the helium balloon. The balloon is designed to counteract most of the weight of the chip encasing, with the drone using the rear propeller to stabilize the altitude. 
                            
\section{Drone Curriculum}
The following is our proposal for a 1-2 week long drone-based curriculum targeting high-school students. We anticipate meeting with students 5 times every week, with each meeting being a normal class-period (e.g., 45 minutes). We describe modules in our lesson plan and discuss possible adaptations and extensions to the curriculum for more advanced students.

\subsection{Programming Fundamentals}
During the first few days, we introduce students to the Python programming language. We start by introducing our custom web-app coding environment. This approach allows students to start experimenting with Python immediately, bypassing the need to install any software for Python development.

We cover the essential Python fundamentals required for the rest of the curriculum, outlined as follows:
\begin{itemize}
    \item Indentation and General Syntax
    \item Dynamic Typing
    \item Functions
    \item Importing libraries and modules
    \item Basic Python data structures (e.g., lists, dictionaries)
    \item Control Flow (e.g., loop syntax, if statements, Python iterables)
\end{itemize}
This module aims to introduce the basics of Python to students with limited programming experience. We expect those with prior coding experience in another language to be able to follow an accelerated timeline, allowing more time for extensions to the final capstone project.

\subsection{Hardware Connection, and Drone API}
Our web-app coding environment is designed to directly interpret Python code and quickly connect to our drones directly so as to abstract the technical details of drone connection away from the curriculum. We begin by presenting a basic example of connecting to the drone and demonstrating simple altitude adjustment. After showcasing the code development process for the drone, we proceed to introduce the application programming interface
(API) documented in Table \ref{tab:drone_control_api}.

\begin{table*}[h]
\centering
\begin{tabular}{lll}
\toprule
\textbf{Function Call} & \textbf{Description} \\
\midrule
\multirow{7}{*}{Drone Movement} & \texttt{up(sec)} & turns on the propellers, moving the drone up for \texttt{sec} seconds \\
& \texttt{down(sec)} & turns on the propellers, moving the drone down for \texttt{sec} seconds \\
& \texttt{off()} & turns off the propellers (except those used for altitude stabilization); makes the blimp stationary \\
& \texttt{turn\_right(sec)} & adjusts the yaw in a clockwise direction for \texttt{sec} seconds \\
& \texttt{turn\_left(sec)} & adjusts the yaw in a counterclockwise direction for \texttt{sec} seconds \\
& \texttt{forward(sec)} & turns on the propellers, moving the drone forward for \texttt{sec} seconds \\
& \texttt{backward(sec)} & turns on the propellers, moving the drone backward for \texttt{sec} seconds \\
\midrule
\multirow{1}{*}{Sensor Data} & \texttt{height()} & queries for the current height of the drone \\
\bottomrule
\end{tabular}
\caption{Drone Control API}
\label{tab:drone_control_api}
\end{table*}

After introducing students to our API, we encourage them to solve a simple task. For example, students may attempt to maneuver the drone through a single suspended hula hoop or retrieve a small weight from a fixed location. With support from the teaching team, this module aims to help students familiarize themselves with algorithmic thinking using physical objects. Through experimentation, students will encounter and address unexpected challenges arising from their chosen tasks.

\subsection{Capstone Project Option 1: Mobile App Integration}
\begin{figure}[h]
    \centering
    \includegraphics[width=0.5\textwidth]{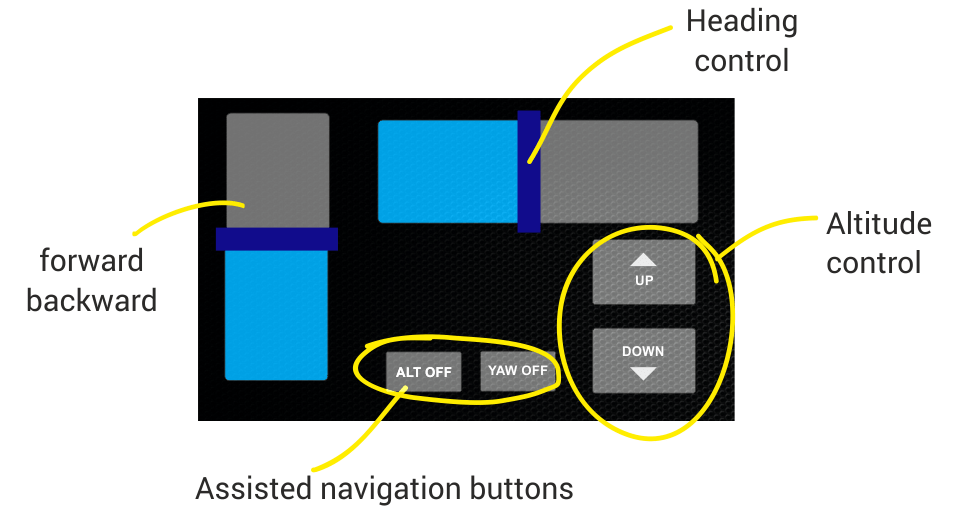}
    \caption{Blimpduino mobile control application \cite{blimpduino_app}}
    \label{fig:blimp_app}
\end{figure}

One of the capstone projects we propose for this curriculum gives students the opportunity to create their own mobile user interface (UI) for drone control. The mobile Control App created by JJRobots is shown in Figure \ref{fig:blimp_app}.

During the beginning of this module, we guide students through setting up XCode for iOS mobile development. We introduce the basics of mobile UI to create buttons, slides, frontend details, and button callbacks. With a brief survey of mobile UI components, students should be equipped to develop their own basic UI to control the drones from their iOS device. We encourage students to then create their own UI components; some possibilities might include a button that triggers a certain movement pattern, or UI components that allow for more sophisticated drone movement.

\subsection{Capstone Project Option 2: Complex Flocking Mechanics}
Another potential capstone project for this curriculum is the experimentation with inter-drone communication. This module encourages more advanced students to study internet communication protocols that can be used to send packets between drones. Once students are able to implement this inter-drone communication, they can take on flocking tasks, where multiple drones coordinate their movements in a synchronized manner, or even implement collision avoidance mechanisms.

\subsection{Final Showcase}
At the conclusion of the curriculum, we organize a final showcase where students share their work. For those that create a mobile UI for drone control, the teaching team will design an obstacle course for students to race against each other. Those that explore drone flocking can exhibit their proposed features with the drones.

\section{Conclusion}
In this paper, we presented a new drone-based curriculum designed to introduce high-school students to computer science principles through hands-on learning experiences. By using inexpensive helium-balloon drones and a custom web-based coding environment, we aim to make programming accessible and engaging for students with varying levels of experience.

Our approach integrates fundamental Python programming skills with practical applications in robotics, encouraging students to solve problems that extend beyond traditional computer programming, and to develop creative solutions. The curriculum includes a range of projects, from basic drone control tasks to advanced capstone projects like mobile app development and inter-drone communication for complex flocking mechanics. By providing a stimulating and supportive learning environment, we aim to inspire a diverse group of students to pursue further studies and careers in computer science and related fields.

Future work will involve implementing this curriculum in high school settings and evaluating its impact on student engagement, learning outcomes, and level of interest in STEM disciplines. We will compile an experience report to document the students' learning journeys and the effectiveness of using drones as educational tools. Our ultimate goal is to create an inclusive and scalable model for computer science education that can be adopted by schools worldwide.







\bibliographystyle{ACM-Reference-Format}
\bibliography{sample-base}

\appendix

\end{document}